# Nonlinear terahertz-wave radiation generated by mid-infrared dual-chirped optical parametric amplification femtosecond pulse laser excitation

Kyuki Shibuya[1, 2, 3*], Kouji Nawata[3], Yoshiaki Nakajima[2, 4, 5], Yuxi Fu[6], Eiji J. Takahashi[6], Katsumi Midorikawa[6], Takeshi Yasui[1, 2] Hiroaki Minamide[3]

[1]Graduate School of Technology, Industrial and Social Sciences, The Tokushima University, Japan
[2]JST, ERATO MINOSHIMA Intelligent Optical Synthesizer (IOS) Project, Japan
[3]Tera-Photonics Research Team, RIKEN Center for Advanced Photonics, RIKEN, Japan
[4]Graduate School of Informatics and Engineering, The University of Electro-Communications, Japan
[5]Department of Physics, Faculty of Sciences, Toho University, Japan
[6]Attosecond Science Research Team, Extreme Photonics Research Group, RIKEN Center for Advanced Photonics, RIKEN, Japan
*Corresponding author: shibuya.tokudai@gmail.com

**Abstract:** We demonstrate an efficient procedure for terahertz-wave radiation generation by applying a nonlinear wavelength conversion approach using a mid-infrared pump source. We used a 3.3 µm dual-chirped optical parametric amplification source and a 1.5 µm fiber laser for a comparison of energy conversion efficiencies. Nonlinear inorganic and organic crystals were used, and for the periodically-poled lithium niobate crystal, the efficiencies were $1.3 \times 10^{-4}$ and $5.6 \times 10^{-12}$ for the 3.3 µm source and the 1.5 µm source, respectively. We confirmed that nonlinear crystals could be pumped with tera-watt / cm$^2$ class by using an mid-infrared source, which reduces several undesirable nonlinear optical effects.

Intense terahertz (THz) wave generation, based on the nonlinear wavelength conversion approach, has revealed various physical phenomena in the THz region, such as the Higgs mode in superconductors, nanoscale crystal growth, and shockwave generation [1–3]. Nonlinear crystals (NLCs) [4,5], gases [6], and liquids [7] have been studied as potential sources for generating intense THz-wave radiation. THz-wave sources, which are based on nonlinear wavelength conversion, generally have high coherency and frequency tunability. Additionally, such THz-wave sources can easily achieve a high gain by considering the phase-matching condition. In particular, lithium niobate (LN) or organic NLCs [8–11] are representative NLCs owing to their large second-order susceptibility, e.g., LN = 25 pm/V (d33) [12], 4-dimethylamino-N-methyl-4-stilbazolium tosylate (DAST) = 210–600 pm/V (d11) [13,14]. NLCs can effectively convert the wavelength by manifesting the phase-matching condition and using an near-infrared (NIR) pump source, whose wavelength corresponds to the low-absorption band in the NLCs. Therefore, an efficient conversion (more than 1%) can be achieved. Further, the electric field strength and energy of the pulse can reach up to MV/cm and mJ order, respectively.

For a more effective nonlinear wavelength conversion, we converted the pump light to THz-wave by using an intense pump source whose intensity reached the terawatt (TW) class. Ti:sapphire and Er-doped fiber lasers are available with a maximum peak power of gigawatt class that can be used as pump sources. These lasers are frequently used because these emission wavelengths of fundamental lights are located in transmission band of NLCs, and they manifest phase-matching conditions, while second harmonic generation (SHG) lights are located in absorption bands of NLCs. Thus, the intensity of pumping has been limited to avoid optical damage of the NLCs by multi-photon absorption (MPA) and absorption of the SHG light [15]. Additionally, the self-focusing effect [16], which causes optical damage to NLCs, appears prominently at shorter wavelengths.

In contrast, THz-wave generation pumped by mid-infrared (MIR) light would enhance the conversion efficiency without optical damage to the NLC. The use of longer wavelength light as a pump source could restrict undesirable nonlinear optical effects and generate a THz-wave with lower pump energy, according to the Manley-Rowe relations [17]. However, despite the knowledge of wavelength conversion using a longer-wavelength excitation source, sufficient experimental studies have not been performed. Based on this hypothesis, we experimentally studied the effectiveness of using MIR pump light for generating THz-waves. However, only a few options were available for high-intensity MIR sources. Recently, a TW class MIR femtosecond (fs) laser has been developed by using the dual-chirped optical parametric amplification (DC-OPA) scheme [18]. The DC-OPA boosts the power up to 210 mJ, 4.2 TW, and it can directly generate two-cycle, carrier-envelope phase-stabilized MIR pulses. These high-intensity MIR sources are expected to efficiently generate THz-waves by reducing the MPA and self-focusing effect. Further, since the longer wavelength light has lower photon energy, it is expected to reduce the rise in temperature of the NLC when the light becomes the dominant absorption band. The DC-OPA laser has a low repetition frequency; therefore, it provides sufficient time for the NLCs to cool down [19]. In contrast, phase-matching condition in the MIR region is not as satisfying as in the NIR region. The absorption of the DAST crystal in the MIR region is higher than that in the NIR region [20–22], and the phonon absorption in NLCs should not be ignored. However, the advantages of the DC-OPA laser will overcome the indicated demerits for the efficient generation of THz-wave radiation.

In this letter, we report THz-wave generation using 3.3 μm and 1.5 μm pump sources. The relationship between the pump intensity and THz-wave output was compared, and inorganic/organic crystals were used as the NLCs; periodically-poled lithium niobate (PPLN), N-benzyl-2-methyl-4-nitroaniline (BNA), and DAST. We present a comprehensive study to reveal the characteristics of the THz-wave radiation generated with several types of NLCs and with different wavelength of pumping sources.

Figure 1 shows the schematic of the experimental setup of our 3.3 μm DC-OPA laser system [18], which consists of two amplification stages. The system delivers 21 mJ pulse energy with a pulse duration of 70 fs and a repetition rate of 10 Hz [18]. In comparison to the 3.3 μm experiment, we constructed an all-polarization-maintaining Er-doped fiber laser using a semiconductor saturable-absorber mirror (SESAM) for mode-locking [23]. The output of the fiber laser was amplified to 150 mW by two stages of Er-doped fiber amplifiers (EDFA). The pulse energy was 6.5 nJ at 1.56 μm. The peak intensity was 8.4 kW with a pulse width of 780 fs and repetition frequency of 22.9 MHz. Table I lists the specifications of the pump sources.

We used the following NLCs at room temperature to generate the THz-waves: 5mol% MgO-doped PPLN (period: 21 μm / crystal length: 5 mm, laboratory made, period: 68 μm / crystal length: 2.7 mm, period: 89 μm / crystal length: 2 mm, HCPhotonics Corp.), DAST crystals (thickness [μm]: 100, 500, 900, 2100, 2540, laboratory-made), and BNA crystals (thickness [μm]: 100, 500, 870, 1330, 1890, laboratory-made). These crystals had no surface coatings. In the case of 3.3 μm pumping, the generated THz-wave was detected using a pyroelectric detector (JASCO Corp.). In contrast, the THz-wave radiation with 1.5 μm pumping was detected by a Fermi-level managed barrier diode (FMB diode, NTT Electronics Corp.) owing to its lower noise equivalent power than that of the pyroelectric detector. These setups were composed of a THz lens for focusing the THz-waves and several black-polyethylene sheets for absorbing the pump light. The temporal electric fields were also measured by a time-domain spectroscopic (TDS) system with the NLCs functioning as emitter and a low-temperature-grown gallium arsenide (LT-GaAs) photoconductive antenna (PCA, G10620-11, Hamamatsu Photonics) was employed as the receiver. The estimated pulse energy along with the measured peak power, temporal width, and pulse repetition rate were calibrated using a pyroelectric detector.

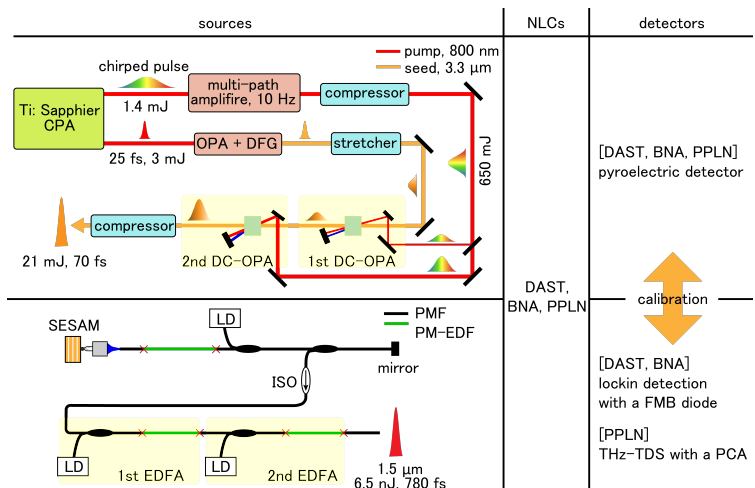

Fig. 1. Illustration of the experimental setup.

Table 1. Specifications of the pump sources.

|  | center wavelength [μm] | frep [Hz] | pulse energy [J] | pulse width [fs] |
|---|---|---|---|---|
| DC-OPA laser | 3.3 | 10 | $21 \times 10^{-3}$ | 70 |
| Er-fiber laser | 1.56 | $22.9 \times 10^6$ | $6.5 \times 10^{-9}$ | 780 |

Figure 2 shows the input/output characteristics of the measured THz-waves pumped at 1.5 μm and 3.3 μm. Considering the phase-matching length and group-velocity dispersion at 3.3 μm pumping, we used a 500 μm thick DAST crystal for generating the THz-waves. The range of pump peak intensity, which was controlled by a variable aperture, was 90 GW/cm$^2$ to 613 GW/cm$^2$ (0.2 mJ to 1.35 mJ) with a beam diameter φ of 2 mm, and maximum THz-wave output of 50 nJ was achieved. In the case of the PPLN (period: 89 μm, phase-matching frequency (backward): 0.49 THz [24]), ϕ was set at 0.5 mm, which corresponded to the window size of the crystal. The pump peak intensity reached 18 TW/cm$^2$, according to the calculation, and the maximum THz-wave output reached 90 nJ (energy conversion efficiency: $1.3 \times 10^{-4}$). Also, in the case of 1.5 μm pumping, PPLN (period: 68 μm) was used in figure 2 and S1, and THz-wave spectra are shown in figure S1.

In both the crystals, the outputs were proportional to the square of the pump energy. This behavior can be explained by the principle of difference frequency generation and optical rectification. The result showed that the second-order nonlinear optical coefficient was almost constant near the MIR region. In particular, the nonlinear optical coefficient had no wavelength dispersion over the 1.5 μm region. In contrast, the phonon-polariton mode of LN crystal has caused multiple absorption bands, these bands change according to pumping wavelength [25]. Thus, the appearance of a typical absorption band should be taken into account when generation of specific frequency was needed. The experiment revealed that the optical damage threshold for 3.3 μm pumping was twice as high as that for 800 nm pumping, as depicted at the bottom of Figure 2. In the case of DAST, this behavior was reasonable while using the 3.3 μm DC-OPA because the repetition rate of less than 50 Hz and ultra-short pulses result in a high damage threshold owing to the reduced thermal effect [19]. Further, saturation of the THz-wave output was not observed in this experiment and no optical damage occurred on both the NLCs.

It should be noted that the gradient of the output curve became larger thus the output itself showed a slight increase for the pumping PPLN at 18 TW, as depicted in the inset of Figure 2; this was probably due to some cascade process. Hauri et al. reported the occurrence of a cascade process in DAST crystals [26]. However, we did not observe such a divergence [26]. In our experiment, the conversion efficiency of DAST was of the order of $10^{-5}$, which also included the loss by the lens and black-polyethylene sheets. This might be caused by a phase-mismatch of the high-frequency THz-waves, which was induced by lack of coherent length when using thick crystals. Therefore, there is a possibility of higher THz-wave output by using thin crystals, whose thickness is less than 100 μm.

For generating intense THz-waves, it is critical to consider the crystal composition and pumping conditions. In the case of PPLN, the LN crystal is damaged by an increased infrared absorption due to SHG [27]. The damage threshold of LN depends on the pulse width, wavelength of the pump source, and amount of MgO doping. Additionally, the thermal damage will increase if the pulse width is large [28]. Therefore, in fs pulse pumping, the damage threshold of the LN crystal without MgO doping is reported to be 10 TW/cm$^2$ [28–30]. However, we confirmed that the THz-wave output increases when the pump energy is increased to 18 TW/cm$^2$. The main reason for this increase was the difference in damage threshold for each pump wavelength [28], and using MgO-doping LN crystal [31]. The maximum pumping energy density for DAST crystals reached 613 GW/cm$^2$. These results could not be achieved with NIR pumping.

Figure 3 shows a THz-wave output dependency of the pulse width at 3.3 μm pumping. The period of PPLN was 21 μm, and phase-matching frequency is 1.7 THz (backward). Here, the pump energy was fixed at 0.7 mJ, and the pulse width was increased by adding several sapphire and CaF2 crystals of different thicknesses. Subsequently, the energy conversion efficiency decreased exponentially as the pulse width increased. Using a short pulse for pump light is an excellent method, however, as Nawata et al. demonstrated, the backward parametric oscillation in the THz region (Bw-TPO) using sub-nanosecond pulse could achieve high-power THz-wave generation [33]. In our experiment, the oscillation was not observed, although the pump peak intensity was higher than 1.6 GW/cm$^2$, which is the threshold of the Bw-TPO. The reason for the absence of the oscillations might be the low temporal interaction between pump light and generated THz-wave in the counter-propagating waves, which is caused by the fs pulse width.

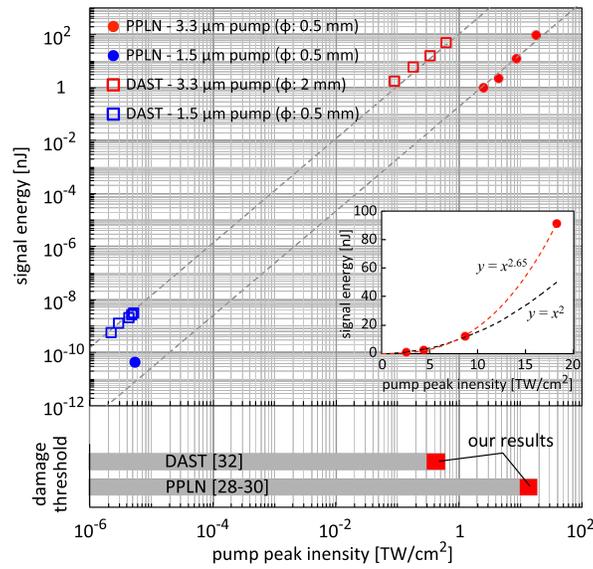

Fig. 2. Comparison of the THz-wave output with pumping at 3.3 μm and 1.5 μm. The gray dotted lines show square curves, and an inset shows the details of PPLN with 3.3 μm pumping.

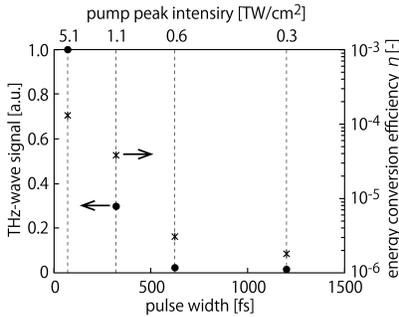

Fig. 3. Experimental results of the THz-wave signal in the 3.3 μm pumping PPLN using several sapphire and CaF$_2$ crystals of different thicknesses.

In an organic crystal, the thickness of the NLC is an important parameter because it affects nonlinear wavelength conversion efficiency [26,34]. The crystal length is also important for actual usage from the perspective of coherence length and group-velocity dispersion. Figure 4 shows the energy conversion efficiency of the DAST crystal. Figure 4 (a) shows the case of 3.3 μm pumping. The pumping intensity was 600 GW/cm$^2$ with the following conditions: 3 mJ, 70 fs, and φ: 3 mm. We observed a decrease in the conversion efficiency when a crystal of thickness greater than 500 μm was used. This decrease in the conversion efficiency was caused by phase-mismatch in the crystal owing to the absence of optical amplification in a thick crystal. This occurs because the pumping wavelength from the phase-matching condition of DAST was highly out of range. Figure 4 (b) shows the energy conversion efficiency at 1.5 μm pumping. The pumping intensity was 10 MW/cm$^2$ under the following conditions: 6.8 nJ, 500 fs, and φ: 0.5 mm. The conversion efficiency increased proportionally with the crystal thickness. This was observed because the generated THz-wave was limited to a low-frequency region (below 1 THz) owing to the narrow bandwidth of the pumping light.

In our experimental setup, we used a THz lens and several black-polyethylene sheets to focus the detector and filter the pump light to emit 3.3 μm wavelength light, respectively. Therefore, the detected THz-wave intensity reduced to approximately 20%. The actual energy of the generated THz-wave was estimated to be several times higher than the observed value. Hence, our result was similar to the outcomes reported in a previous paper [15].

In conclusion, we generated THz-wave using a TW class MIR ultra-short pulse source for overcoming the pumping limitation of NLCs. The experiment demonstrated an improvement in the damage threshold value to 613 GW/cm$^2$ in the case of organic NLCs and to 18 TW/cm$^2$ for inorganic NLCs. We could successfully avoid undesirable nonlinear optical effects. The DC-OPA laser enabled pumping of the NLCs with extremely high energy, and as a result, efficient THz-wave generation was achieved. As further discussed in Supplement.

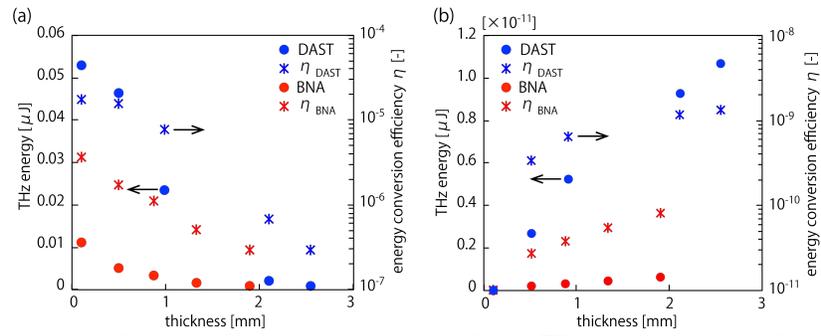

Fig. 4. Relationship between the thickness of organic nonlinear crystals and THz-wave output. (a) 3.3 μm DC-OPA pumping. (b) 1.5 μm Er-doped fiber laser pumping.

The outcomes of our study can be useful for generating intense THz-waves using the nonlinear wavelength conversion method. Intense THz-wave generation using organic NLCs will be an attractive option because of their large second-order susceptibility, possibility of large area growth, printability, and flexibility [35]. Furthermore, energy scaling would be a useful approach for increasing the THz-wave output without causing damage to the crystal. The total excitation energy can be improved by expanding the beam diameter and using a larger area of the NLCs while maintaining the energy density of the pump light. Based on energy scaling, we estimated that the output would be more than 470 μJ and 15 μJ, if we use a 1 $cm^2$ PPLN and DAST crystal, respectively, without the optical component induced losses. These exceptional capabilities are useful for commercial and large-scale applications of THz-wave radiation.

**Funding.** Japan Society for the Promotion of Science (JP17H01282, JP18H01906); Exploratory Research for Advanced Technology (JPMJER1304).

**Acknowledgments.** The authors thank all the members of the Tera-Photonics Laboratory at RIKEN for their helpful cooperation. The authors also appreciate Mr. K. Nitta of Tokushima University for his help in the preparation of the THz-TDS system.

**Disclosures.** The authors declare no conflicts of interest.

**Data availability.** Data underlying the results presented in this paper are not publicly available at this time but may be obtained from the authors upon reasonable request.